\documentclass[twocolumn,notitlepage,amsmath,amssymb,showpacs,showkeys,floatfix,prl]{revtex4-1}
\usepackage[utf8]{inputenc}
\usepackage{amsmath}
\usepackage{graphicx}  
\usepackage{tabularx}
\usepackage{appendix}

\DeclareMathOperator*{\argmax}{arg\,max}
\begin{document}
\title{Theoretical guidelines for editing ecological communities}
\author{Vu Nguyen, Dervis Can Vural}
\affiliation{University of Notre Dame, South Bend, IN}
\date{\today}
\begin{abstract}
Having control over species abundances and community resilience is of great interest for experimental, agricultural, industrial and conservation purposes. Here, we theoretically explore the possibility of manipulating ecological communities by modifying pairwise interactions. Specifically, we establish which interaction values should be modified, and by how much, in order to alter the composition or resilience of a community towards a favorable direction. While doing so, we also take into account the experimental difficulties in making such modifications by including in our optimization process, a cost parameter, which penalizes large modifications. In addition to prescribing what changes should be made to interspecies interactions given some modification cost, our approach also serves to establish the limits of community control, i.e. how well can one approach an ecological goal at best, even when not constrained by cost.
\end{abstract}
\maketitle
\section{Introduction}
Controlling ecological communities has so far had mixed success. Since the population dynamics of a community can sensitively depend on the precise values of interactions and species abundances, attempts motivated by qualitative reasoning has lead to ineffective control of targeted species or adverse outcomes on untargeted species \cite{hoddle,louda,messing,schlaepfer,vandriesche}. An engineer can design a complex electronic device on the drawing board before building a fully functioning prototype. Can an ecologist design communities in a similar way? Our aim here is to develop analytical methods that might serve as a guide for manipulating the composition and resilience of communities.

In the literature we see two types of problems that motivate such quantitative approaches. The first concerns with eradicating invasive species and pests, typically by releasing predators. These studies are limited to few-species or few trophic levels such as prey and predator \cite{jiang,jiang2,liu,liu2,Zhang2016}, prey, predator and super predator \cite{baek}, one predator and multiple prey \cite{georgescu},  one prey and multiple predators \cite{pei}, and two prey and two predators \cite{Rafikov2008}. These control schemes employ feedback response, which requires frequent measurements of abundances, which is challenging even for small communities.

The second concerns controlling species abundances in a chemostat in order to maximize the production of various useful proteins \cite{Yang2019,Sun2011,Wei2013,Cheng2012,Leenheer2003}. These are consumer resource models which treat resource influx and dilution rates as control parameters. While it is much more feasible to monitor species abundances and implement precise control curves in a chemostat, these communities are also relatively simple, since in such uniform environments the principle of competitive exclusion eliminates all but few of the species \cite{Sommer1983,Xu2016}. More sophisticated chemostat models have been proposed \cite{Mazenc2009,Mazenc2012} which restricts the growth rates or introduces additional nutrients in order to sustain multiple species.

An ideal ecological control theory must be scalable: the control procedure should not sensitively depend on the size or the complexity of the community. It should work for a diverse variety of trophic structures. It should also be experimentally feasible: it should not require harvesting or breeding control species in real time according to precise curves, or require continuous monitoring the abundances of many species.

\begin{figure}
\includegraphics[width=1.01\linewidth]{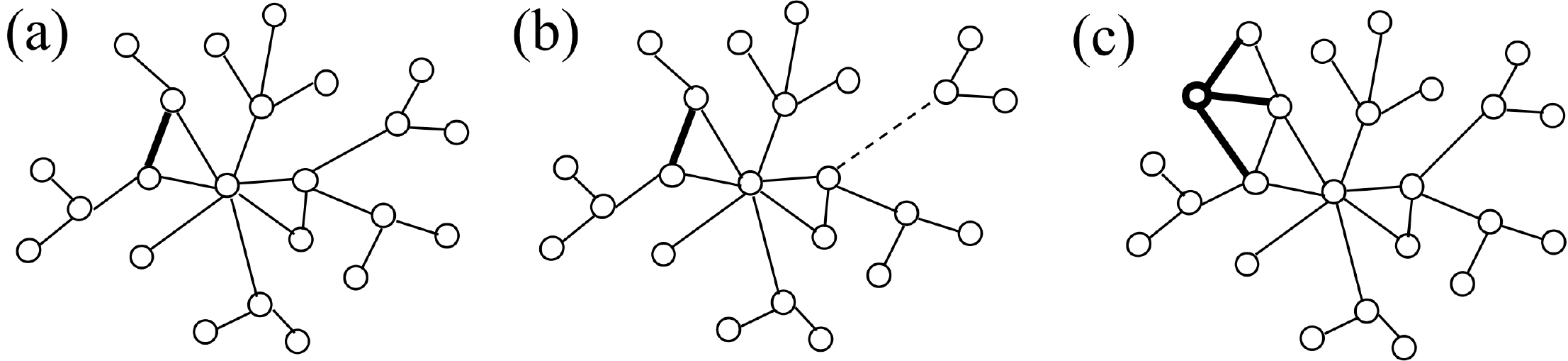}
\caption{{\bf Three control schemes for editing ecological communities.} (a) Modifying an interaction (bold) to displace equilibrium abundances. (b) Modifying an interaction (bold) to reduce displacements caused by a random latter change (dashed). (c) Adding an exogenous species with tailored interactions (bold) to displace equilibrium abundances.}
\label{schematics}
\end{figure}

Here we explore the possibility of making a one-time change in an ecological community in order to shift its equilibrium composition towards a desirable target, or increase its resilience. Our framework consists of identifying the ideal attributes of a ``control species'', which, when introduced into the community once, will permanently alter its composition or stability.

Experimentally, there are multiple ways to realize changes in the interaction matrix. Genes and transcription factors that modulate interspecies interactions can be targeted through gene editing \cite{TetardJones2007,Mooney2008,Broekgaarden2008,Whitham2012,AbdalaRoberts2012,Lamit2015}. If the strain with the new interaction is more fit, it will naturally invade. If not, the original strain must be annihilated first so that the engineered strain can fill its niche. Second, one could make  use  of  the  diversity  of  interaction  values  already present in a population (for example, some predators may already be better at catching prey) \cite{Mooney2011,Barker2018,Zytynska2019}. In this case, one would isolate an individual with the desirable interaction value, culture it, and add it back to the original community in much larger numbers. Interaction values can also be modified by environmental factors such as temperature, pH, and chemicals \cite{Tylianakis2008,Englund2011,Rall2012,Sentis2012,Griffiths2015,Ratzke2018,Mugabo2019,Niehaus2019}. For example one could use a drug that targets a protein responsible for mediating a particular interspecies interaction.





Despite these possibilities, engineering species with desired interaction properties presents many technical challenges. As interesting as these challenges are, here we fully omit the problem of actually \emph{building} control species and instead focus on the problem of \emph{designing} control species, --just as an engineer might overlook how circuit elements are actually manufactured and focus on the design of a circuit. Furthermore, our goal is to determine the characteristics of the \emph{ideal} control species, even though a real species, natural or engineered, might be a mere approximation of this ideal, -- just as real circuit elements are approximations of ideal ones.

{\bf Problem Statement.} We will work with the standard Lotka-Volterra equations, which describe the population dynamics of sparse, well-mixed communities
\begin{align}
\dot{n}_i= n_i\bigg(r_i+\sum_{j=1}^{n}A_{ij} n_j\bigg)\label{eq:lv}
\end{align}
where the abundance $n_i(t)$ of species $i$ changes according to its intrinsic growth rate $r_i$ and its interactions with others $A_{ij}$.

Our goal is to solve the following three problems. {\bf (1)} Modify an interaction matrix element such that the equilibrium community composition shifts towards a new desirable state. {\bf (2)} Modify an interaction matrix element so that a latter random change influences the community composition minimally. {\bf (3)} Add a novel species (a new row and column to the interaction matrix) to shift the community composition towards a new desirable state. While addressing these problems, we will also establish theoretical limits to community control: we will determine how much one can vary the composition and resilience of a community at best, even if one could introduce arbitrarily large changes to the interaction values.

The coexistent equilibrium of Eqn.(\ref{eq:lv}), $\vec{n}=\vec{x}=-{\bf R}\vec{r}$, is obtained by setting the parenthesis to zero  (provided that ${\bf R}\equiv {\bf A^{-1}}$ exists and $x_i>0$). For our purposes, we assume that our system originally resides in a coexistent state in stable equilibrium. As we will see, our control protocols will largely succeed in maintaining both the stability and coexistence, however this is not guaranteed.

\section{Results}
{\bf Control scheme 1: Interaction modification for displacing equilibrium abundances.} 
Our first control scheme involves modifying the interaction between two species in order to move the original equilibrium composition $\vec{x}$ as close as possible to a desirable target $\vec{y}$. 



A change in a single matrix element $A_{ab}\rightarrow A_{ab}+\epsilon$ leads to a change in its inverse ${\bf R\to R'}$. We use the Woodbury matrix identity to find
\begin{align}
    R_{ij}'= R_{ij}-\epsilon \frac{R_{ia}R_{bj}}{1+\epsilon R_{ba}}.\label{inv}
\end{align}
Upon modifying the matrix element, the equilibrium composition becomes
\begin{equation*}
    \vec{z}=-{\bf R'}\vec{r}=\vec{x}-\mu\vec{w}_a
\end{equation*}
where $\vec{w}_a$ is the $a^\mathrm{th}$ column of ${\bf R}$ and $\mu=\epsilon x_b / (1+\epsilon R_{ba})$. This tells us that the equilibrium can be displaced along only certain directions $\vec{w}_a$, indexed by $a=1,2,...,N$. If the desired target is along one of these directions then we can hope to reach it precisely. Otherwise, the best we can do is to come close. 

\begin{figure}
\centering
\includegraphics[width=\linewidth]{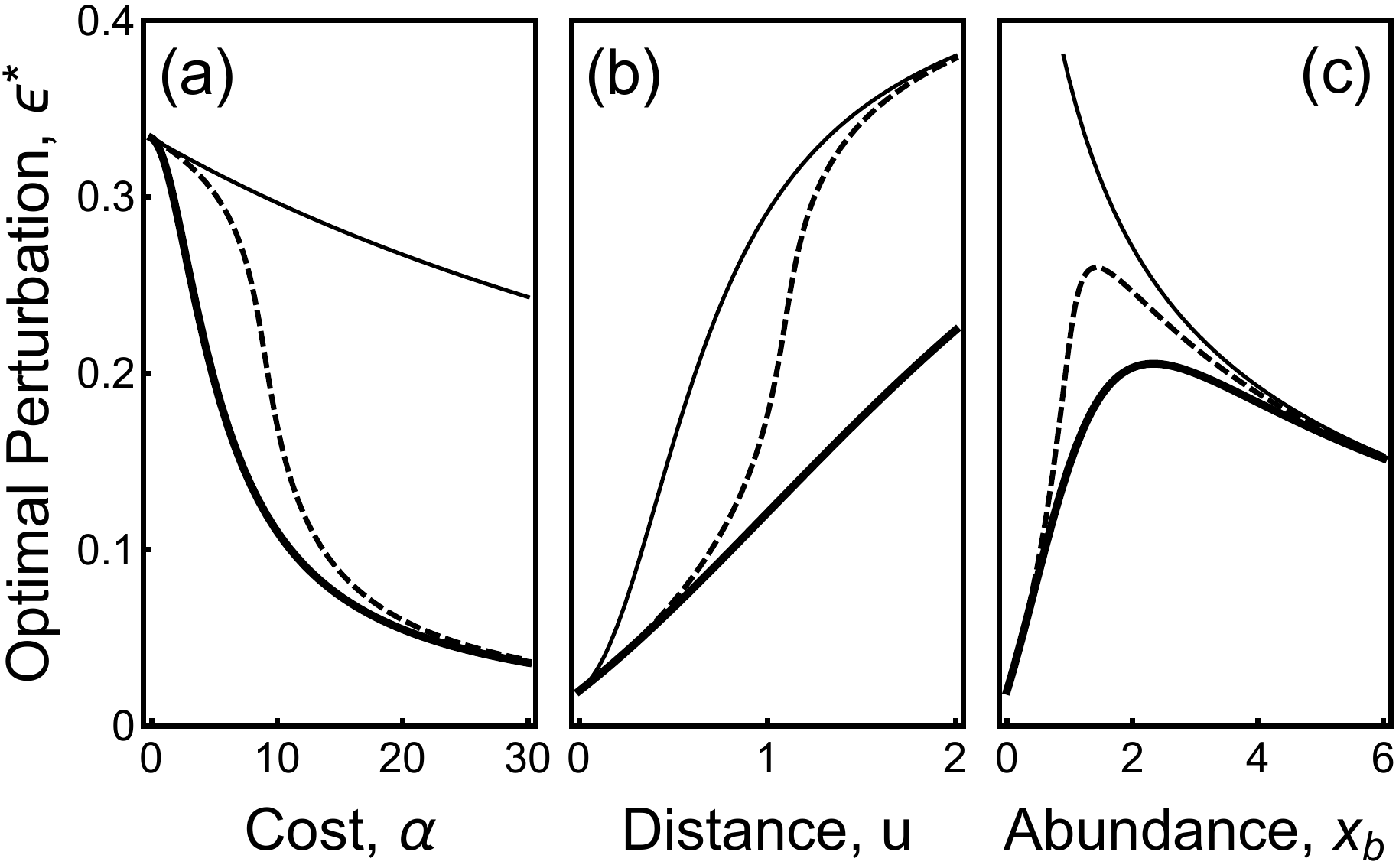}
\caption{{\bf Control scheme 1: Optimal perturbation to a single matrix element $A_{ab}$.} The exact analytical solution to Eqn.(\ref{epsstar}) for the global minimum of $\mathcal{L}(\epsilon)$ (dashed) and simple asymptotic forms Eqn.(\ref{eq:single_result}) (thin line) and Eqn.(\ref{smallalpha}) (thick line) valid for large and small values of $\alpha$ respectively. Parameter values are $w_a^2=1, x_b=1,\vec{u}\cdot\vec{w}_a=-1, R_{ba}=-2$ and $\alpha=10$ unless one is varied in the horizontal axes.}
\label{fig2}
\end{figure}

We will find which interaction matrix element should be modified, and by how much, by minimizing
\begin{equation}
    \mathcal{L}=|\vec{y}-\vec{z}|^2+\alpha \epsilon^2.
    \label{eq:cost_function_sweep}
\end{equation}

The first term ensures that we come close to our target, while the second term accounts for the difficulty in making changes to interspecies interactions. $\alpha$, the only adjustable parameter in this letter, is the modification cost per modification. It quantifies the relative importance of making small changes versus approaching our target. The smaller the cost, the larger our modification can be, and the closer we can approach our target.

We should caution that for sufficiently large perturbations, $\epsilon\to-1/R_{ba}$, Eqn.(\ref{inv}) blows up. Such changes will destabilize the community and collapse it into a smaller community after a cascade of species extinctions. Dealing with this singularity is mathematically challenging, and for the most part practically undesirable, so most analytical results here concern with perturbations that are far away from this singularity. Specifically, we will focus on the ``large'' $\alpha$ regime (leading to small perturbations that fall short of hitting the singularity) and the small $\alpha$ regime (leading to large perturbations pushing us beyond the singularity).

\begin{figure}
\centering
\includegraphics[width=\linewidth]{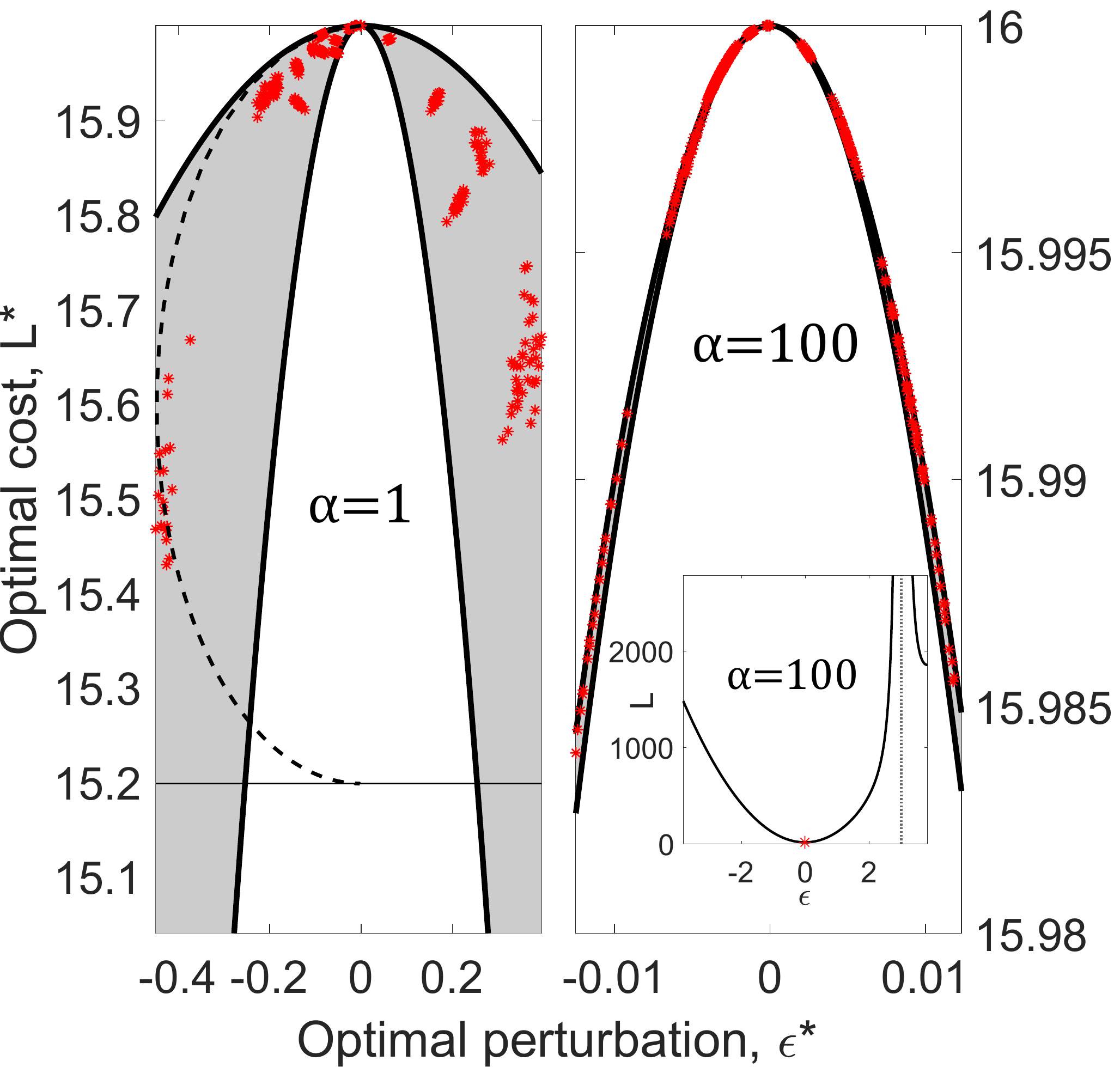}
\caption{{\bf Control scheme 1: Optimal perturbation to a single $A_{ab}$ for all $a,b$.} Each red dot is the global minimum of $\mathcal{L}$ as obtained from the exact analytical solution to Eqn.(\ref{epsstar}) for a given $(a,b)$ pair. Eqn.(\ref{eq:L_opt_quadratic}) is used to sweep across pairs $(a,b)$ for the $n^2$ parabolas (grey), which are bound between two special parabolas (black). The thin horizontal line shows the limit as $x\rightarrow\infty$ for the best $\vec{w}_a$. The inset shows the cost function for the best $(a,b)$ with the singularity marked by a vertical line. For system parameters cf. Methods.}
\label{fig:opt_multi}
\end{figure}

\begin{figure}
\centering
\includegraphics[width=\linewidth]{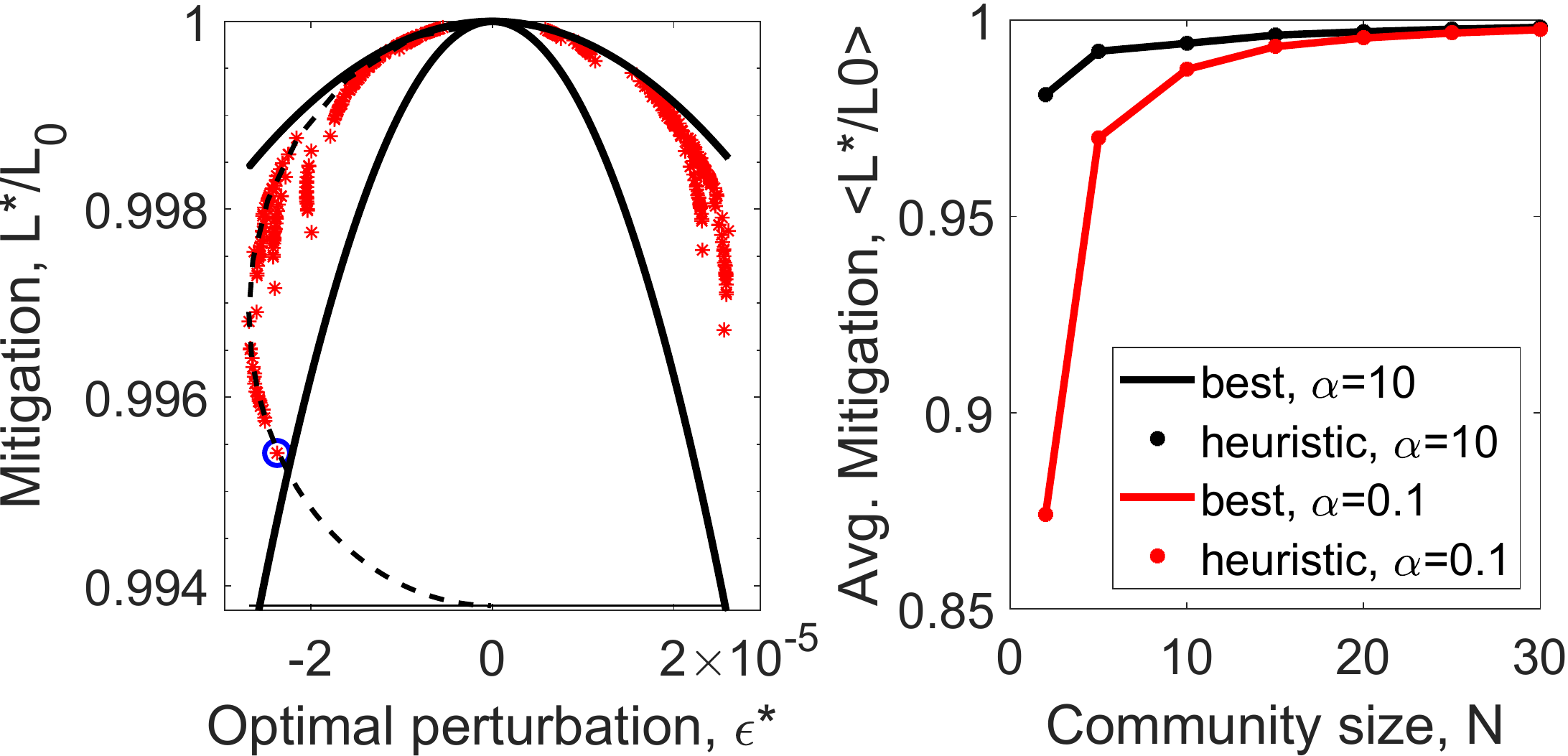}
\caption{{\bf Control scheme 2: Optimal preventive modifications.} We plot Eqn.(\ref{eq:prevent}) for all $(a,b)$ pairs (left) for $\alpha=10$. The blue circle is the $(a,b)$ for which $x_b=x_{\max}$ and  $a=\argmax_a \vec{w}_a\cdot\langle\vec{w}\rangle$ and indeed minimizes $\mathcal{L}$. The solid black bounding lines are based on Eqn.(\ref{eq:L_opt_prevent}) for $x_b^2\vec{w}_a^2\rightarrow 0$ and $x_{\max}^2w^2_{\max}$. The dashed ellipse is a sweep varying $x$ for the ``best'' $\vec{w}_a$ in Eqn.(\ref{eq:prevent}) and Eqn.(\ref{eq:L_opt_prevent}) on normalized coordinates. The thin horizontal limit shows the limit as the $x\rightarrow\infty$ for the best $\vec{w}_a$. We compare the average performance (right) for different community sizes using our heuristic method (dots) against the best performance found by testing all possible cases (lines). For system parameters cf. Methods.}
\label{fig:prevent_opt_sweep}
\end{figure}

To minimize $\mathcal{L}$ we solve for $\epsilon^*$ that satisfies $\left.\frac{d\mathcal{L}}{d\epsilon}\right|_{\epsilon^*}=0$. Taking the derivative and reorganizing terms, we get 
\begin{align}
    \epsilon^*&=\gamma/\left[\alpha(1+\epsilon^* R_{ba})^3-\beta\right]\label{epsstar}\\
    \gamma&=-x_b\vec{u}\cdot \vec{w}_a,\quad \beta=R_{ba}\gamma-x_b^2|\vec{w}_a|^2,\quad \vec{u} = \vec{y}-\vec{x}.\nonumber
\end{align}
Eqn.(\ref{epsstar}) is a quartic polynomial with an exact analytical solution shown in Fig.\ref{fig2}. Since the analytical formula is rather lengthy, we omit it here and simply refer to \cite{weisstein}. 

Now we focus on the special cases of large and small $\alpha$, which will be useful as we move on, as well as biologically more insightful. We start by observing that Eqn.(\ref{epsstar}) has $\epsilon^*$ both on the right and left side, which we arranged in this way to solve for $\epsilon^*$ perturbatively. For $\alpha\to\infty$, we have $\epsilon^*=0$. For large but finite costs, we can plug in $\epsilon^*=0$ to the right, and get $\epsilon^*=-\gamma/(\beta-\alpha)$ on the left. Then we repeat, plugging this into the right, to get
\begin{align}
\epsilon^*\simeq \frac{\gamma}{\alpha[1-\gamma R_{ba}/(\beta-\alpha)]^3-\beta}\qquad\mbox{(Large } \alpha \mbox{ limit)}\label{eq:single_result}
\end{align}
on the left. This procedure very rapidly converges to the true solution, and our numerical tests show that Eqn.(\ref{eq:single_result}) is already quite accurate for practical purposes. 

We follow a similar procedure for small costs. For $\alpha\to0$, Eqn.(\ref{epsstar}) gives $\epsilon^*=-\gamma/\beta$. For small but non-zero costs, this solution can be plugged back to the right side of Eqn.(\ref{epsstar}), 
\begin{align}
\epsilon^*\simeq\frac{\gamma}{\alpha(1-\gamma R_{ba}/\beta)^3-\beta}\qquad\mbox{(Small } \alpha \mbox{ limit)}\label{smallalpha}
\end{align}

Note that the second order approximations Eqns.(\ref{eq:single_result}) and (\ref{smallalpha}), as well as their first order analogues overlap for $\alpha\to0$ and $\alpha\to\infty$, but depart for intermediate values. Fig.\ref{fig2} compares these asymptotic forms with the exact analytical solution of Eqn.(\ref{epsstar}). In all panels, we take the root of the equation corresponding to the global minimum of the quartic polynomial.

Eqns.(\ref{eq:single_result}) and (\ref{smallalpha}) prescribe the ``best bang for the buck'' amount of change that must be introduced to $A_{ab}$ for a \emph{particular} pair $(a,b)$, for large and small costs. But there are $N^2$ pairwise interactions, and thus $N^2$ optimal $\epsilon^*$'s. Which $(a,b)$ pair is the best to modify? 

Ideally, we should substitute $\epsilon^*$ into Eqn.(\ref{eq:cost_function_sweep}) for all $(a,b)$, and identify the $(a,b)$ that minimizes $\mathcal{L}$. However, we outline an easier and more insightful way to obtain the best $(a,b)$ for small and large costs. We start by writing Eqn.(\ref{eq:cost_function_sweep}) as,
\begin{equation}
    \mathcal{L}(\epsilon)= u^2+\frac{x_b^2|\vec{w}_a|^2 \epsilon^{2}}{(1+\epsilon R_{ba})^2} - \frac{ 2\epsilon \gamma}{1+\epsilon R_{ba}}+\alpha \epsilon^{2}\label{L}
\end{equation}
In the large $\alpha$ limit ($\epsilon^*\to\gamma/(\alpha-\beta)$ and $\alpha\gg \gamma R_{ba}$),
\begin{align}
\mathcal{L}^*=u^2\!-\!\gamma^2/\left(\alpha\!+\!w_a^2 x_b^2\right)=u^2\!-\!x_b^2 (\vec{u}\cdot\vec{w}_a)^2/\left(\alpha\!+\!w_a^2 x_b^2\right)\nonumber
\end{align}
As we see, to minimize this, we must simply pick the $b$ corresponding to the species with largest abundance $x_b$ and the $a$ that maximizes the dot product $\vec{u}\cdot\vec{w}_a$. Both of these conditions make biological sense.

The collection of all species interactions $(a,b)$, define a large number of parabolas, plotted gray in Fig.\ref{fig:opt_multi}. Interestingly, these parabolas are bounded above by $\mathcal{L}^*\approx |\vec{u}|^2 \!-\!\alpha \epsilon^{*2}$ and below by $\mathcal{L}^*\approx |\vec{u}|^2 \!-\!(\alpha\!+\!x_\mathrm{max}^2w_\mathrm{max}^2) \epsilon^{*2}$ where $x_\mathrm{max}$ is the largest population abundance and $w^2_\mathrm{max}$ is the displacement vector with the largest norm. These bounds are shown in Fig.\ref{fig:opt_multi} and can be readily obtained by writing $\gamma$ in terms of $\epsilon^*$ using Eqn. (\ref{epsstar}),
\begin{equation*}
    \gamma=\left[\alpha(1+\epsilon^*R_{ba})^3+x_b^2|\vec{w}_a|^2\right]\frac{\epsilon^*}{1+\epsilon^* R_{ba}}
\end{equation*}
and then plugging this into Eqn.(\ref{L}),
\begin{align}
\mathcal{L}^*\approx |\vec{u}|^2 -(\alpha+x_b^2w_a^2)\epsilon^{*2}.
\label{eq:L_opt_quadratic}
\end{align}

Now we turn to the $\alpha\to0$ limit ($\epsilon^*\to-\gamma/\beta)$, where Eqn.(\ref{L}) gives
\begin{align}
\mathcal{L}^*=u^2+\gamma^2 \left(\frac{\alpha}{\beta^2}-\frac{1}{w_a^2x_b^2}\right)\simeq u^2(1-\cos^2\!\theta_a)\nonumber
\end{align}
Here $\theta_a$ is the angle between $\vec{u}$ and $\vec{w}_a$. This tells us that when the controller is not constrained by cost, they should simply pick the displacement vector $\vec{w}_a$ best aligned with the desired displacement direction and then they could modify any element $b$. 

The large-$\alpha$ result emphasized the effectiveness of small modifications which required taking advantage of species with large population abundances and strong displacement magnitudes to more easily propagate our perturbation. In this regime, however, interaction cost does not matter and the only restriction is based on the natural displacement directions of the community.

The inset in Fig.\ref{fig:opt_multi} shows a numerical example where Eqn.(\ref{eq:single_result}) minimizes $\mathcal{L}$. Each red star in the left and right panels of Fig.\ref{fig:opt_multi} is such an optimal solution for different $(a,b)$ pairs. Most interaction modifications do not effectively shift the abundances towards our target $\vec{y}$ but there exists a couple of key interactions which will perform well at a much lower overall cost. 

For this and all following numerical examples we use the parameter values and procedures described in the Methods section.

{\bf Control scheme 2: Interaction modification for minimizing community vulnerability.} 
The interactions between species are mediated by heritable phenotypes, which, like any other trait, are selected upon. Thus, as species adapt to each other, the interaction matrix will change \cite{abrams,abrams2,diekmann,schaffer,dercole,Friesen2004,Valdovinos2010,Smith2015}. In particular, it was shown experimentally that interactions typically change in one direction \cite{fiegna,rivett}.

In this section we address which matrix element we should modify, and by how much, $A_{ab}\to A_{ab}+\epsilon$, in order to minimize the expected equilibrium displacement upon a latter change $A_{cd}\to A_{cd}+\zeta$ on some random $(c,d)$.

Since we cannot know ahead of time where the random mutation may occur, we aim to minimize the displacement averaged over all possible end locations $(c,d)$, while also taking into account the cost of modification as before. To this end, we define a cost function,
\begin{align*}
    \mathcal{L}&=\langle|\vec{y}-\vec{x}|^2\rangle_{cd}+\alpha\epsilon^2
\end{align*}
If we assume that the changes in the interaction matrix are small ($\epsilon R_{ba}\ll 1$ and $\zeta R_{dc}\ll 1$) then we can use the formulation present in the previous section twice. When both shifts are small, the new fixed point of the community can be linearized such that $\vec{y} \simeq \vec{x} - \epsilon x_b \vec{w}_a - \zeta x_d \vec{w}_c$. In this case,
\begin{align}
\mathcal{L}\simeq\epsilon^2 x_b^2 w_a^2+\zeta^2 \langle x^2\rangle\langle w_c^2\rangle_c
    +2\epsilon\zeta x_b\langle x\rangle\langle\vec{w}_a\cdot\vec{w}_c\rangle_{c}+\alpha\epsilon^2,\nonumber
\end{align}
where the first term represents the displacement from the initial fixed point caused by the preventive modification, the second is the expected displacement the initial fixed point, the third  represents the ``preventive'' benefit caused by our modification, and the final term is the cost of the modification. Setting $d\mathcal{L}/d\epsilon=0$ we find the optimal modification if we were to implement it at $(a,b)$
\begin{equation}
    \epsilon^*=-\zeta x_b\langle x\rangle\langle\vec{w}_a\cdot\vec{w}_c\rangle_{c}/(x_b^2w_a^2+\alpha),
    \label{eq:prevent}
\end{equation}
and the optimal cost is given by
\begin{equation}
    \mathcal{L}^*=\mathcal{L}(\epsilon^*)=\zeta^2 \langle x^2\rangle\langle\vec{w}_c^2\rangle_c-( x_b^2 w_a^2+\alpha)\epsilon^{*2}.
    \label{eq:L_opt_prevent}
\end{equation}
Note that in absence of any preventive modification ($\epsilon=0$) the cost function is just the expected average displacement from $\vec{x}$. 

As before, Eqn.(\ref{eq:L_opt_prevent}) is constrained between two parabolas given by substituting $x_b^2 w_a^2\to0$ and $x_b^2w_\mathrm{max} w_\mathrm{max}^2$. These two parabolas are shown in Fig.\ref{fig:prevent_opt_sweep}.

Minimizing the cost across all possible locations $(a,b)$ is equivalent to maximizing the second term
\begin{equation*}
    \max_{ab}\quad ( x_b^2 |\vec{w}_a|^2+\alpha)\epsilon^{*2}=\max_{ab}\quad \frac{\zeta^2 x_b^2\langle\vec{w}_a\cdot\vec{w}_c\rangle^2_{c}}{x_b^2w_a^2+\alpha}.
\end{equation*}
Fixing $a$ gives a monotonically increasing function of $x_b$. Thus again, it is best to pick the $b$ for which $x_b$ is the largest. Then maximizing with respect to $a$ requires 
\begin{equation*}
    a = \argmax_a[\langle\vec{w}_a\cdot\vec{w}_c\rangle_c^2/(x_{\max}^2\vec{w}_a^2+\alpha)]
\end{equation*}
In the large $\alpha$ limit ($\alpha \gg x_b^2w^2_{\max}$) the best $a$ is the one corresponding to the $w_a$ which on average gives the largest dot product with the average $\vec{w}$
\begin{equation*}
    a=\argmax_a \langle\vec{w}_a\cdot\vec{w}_c\rangle_c=\argmax_a [\vec{w}_a\cdot\langle \vec{w}\rangle]
\end{equation*}
In other words, we would like to pick the $\vec{w}$ that has the largest projection onto the average $\vec{w}$. Applying the small $\alpha$ limit ($\alpha\ll x_b^2 w_{\max}^2$) gives the same result. Sweeping across possible values for $x$ in Eqn.(\ref{eq:L_opt_prevent}) for some finite value of $\alpha$ and fixed $\vec{w}_a$ draws an half-ellipse located on the positive or negative side depending on the sign of $\epsilon^*$ as shown by the dashed line in Fig.\ref{fig:prevent_opt_sweep}.

We generate a random interaction matrix and show with red asterisks, the improvement in $\mathcal{L}$ upon modifying different $(a,b)$ pairs in Fig.\ref{fig:prevent_opt_sweep}. The best $(a,b)$ pair, chosen according to the arguments presented above, is marked blue, and indeed minimizes $\mathcal{L}$.

We observe that while the best position which minimizes our cost function does not necessarily provide the best reduction in displacement per $\epsilon$, it is among the best performers. For example, in Fig.\ref{fig:prevent_opt_sweep}, left, minuscule change of $\epsilon^*=-2.4\times10^{-5}$ in an optimally chosen interaction matrix element ($A_{ab}=6.6\times10^{-2}$) can reduce the displacement of abundances by $\sim0.2\%$ for a 30 species community. While the ratio of these numbers are impressive, the absolute scale of change is small. Furthermore, for large communities, this very effective manipulation does not scale up to larger changes: we can improve the resilience of a large community only so much by manipulating only a single interaction. In contrast, we find that smaller communities have higher relative reductions in displacement than larger communities (cf.  Fig.\ref{fig:prevent_opt_sweep}, right)


{\bf Control scheme 3: Introducing an exogenous species for displacing equilibrium.} For our third control scheme we  consider the possibility of choosing an exogenous species from a library of options, with given interactions and then determining how these interactions should be edited as to displace the equilibrium abundances of the community as close as possible to a given target. As before, we will account for the difficulty in making such changes and introduce a cost per change.

\begin{figure}
\centering
\includegraphics[width=\linewidth]{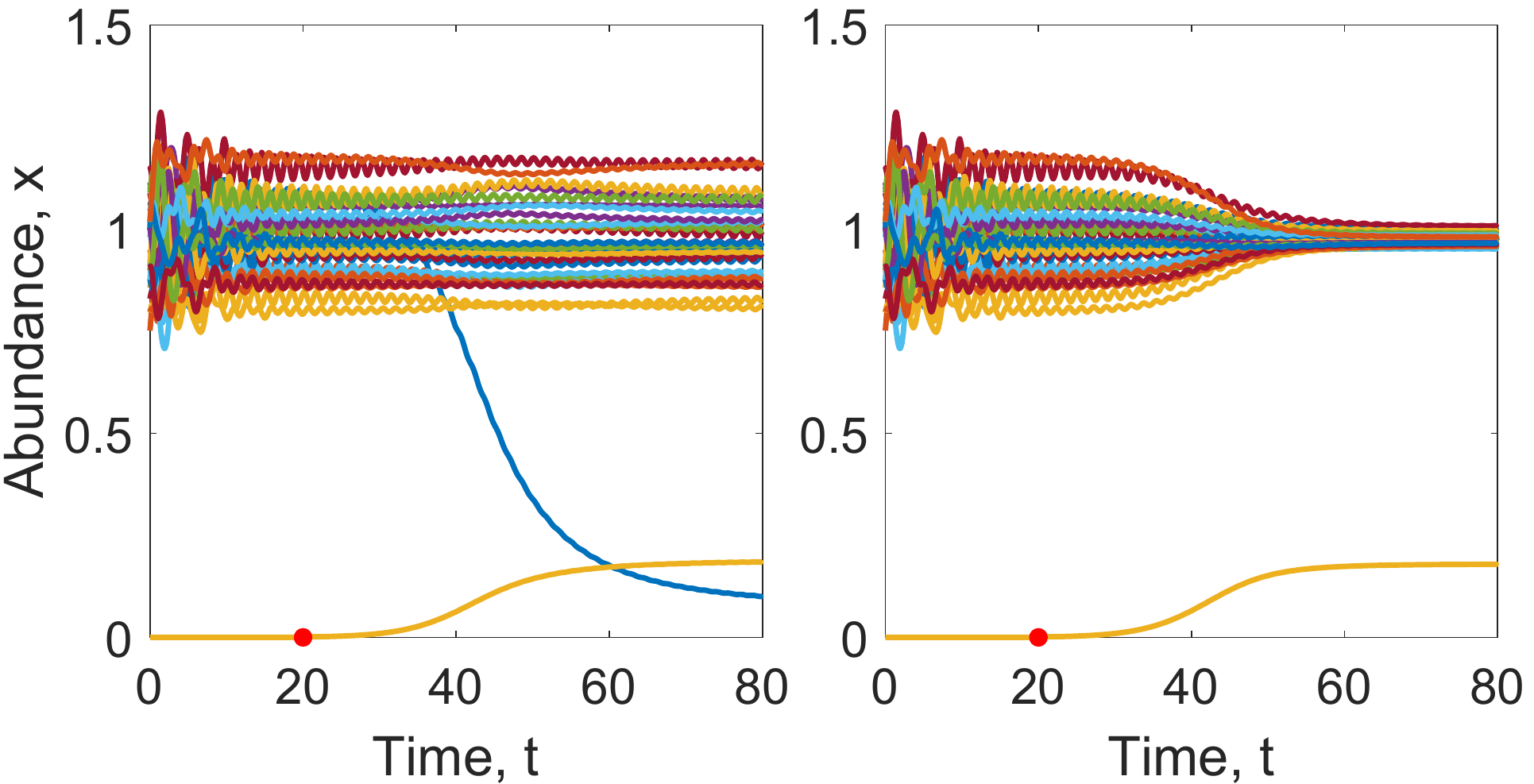}
\caption{{\bf  Control Scheme 3: Introducing edited exogenous species.} To illustrate the effectiveness of our equations we work out two examples where {\bf (left)} we drive a ``pest'' towards extinction while keeping others unperturbed, and {\bf (right)} we equalize all species abundances to $1$. The control species is introduced at $t=20$ (red dot, $\alpha=10^{-3}$) by a very small amount ($10^{-3}$) and even after fixing, is much smaller ($\sim0.1$) than the others, but still able to significantly displace equilibrium. For system parameters cf. Methods}
\label{fig:control_species_vector}
\end{figure}

\begin{figure}
\centering
\includegraphics[width=\linewidth]{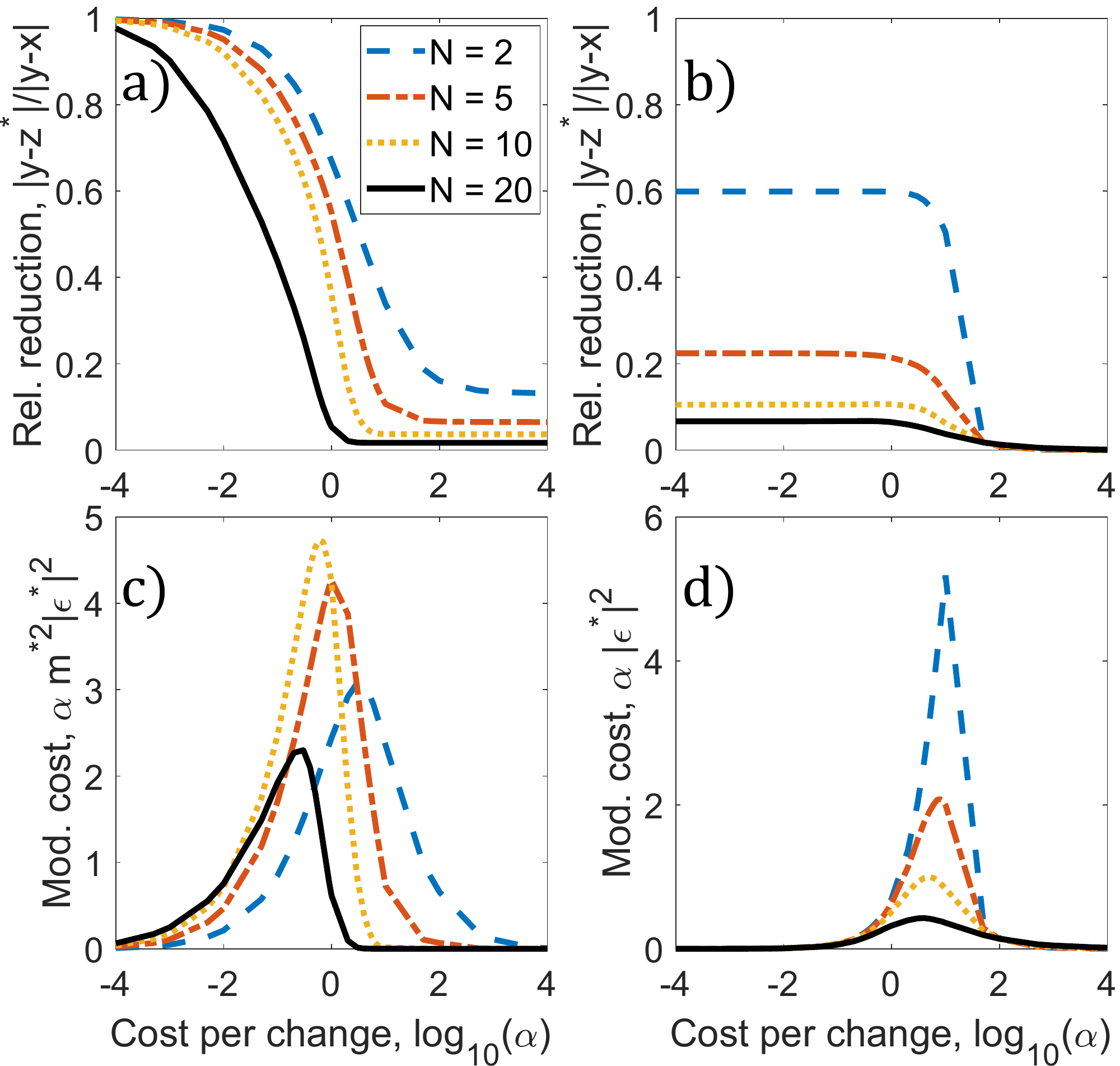}
\caption{ {\bf Economics of species editing.} We generate random communities and shift their equilibrium by using randomly generated (and then systematically edited) exogenous species ({\bf left coloumn}) or by only modifying a single interaction which gives the best result ({\bf right coloumn}). {\bf Top row:} Plotting the average relative reduced distance from the desired target as a function of cost shows an abrupt transition: once the cost of interaction modification is below a critical threshold we are able to introduce the changes necessary to approach our target. {\bf Bottom row:} The average total cost of species editing, $\alpha m^{*2}\left|\vec{\epsilon}^*\right|^2$, for the exogenous species and the single interaction, $\alpha\left|\epsilon^*\right|^2$. For $\alpha\to0$, we reach our target without needing to pay much for scheme 3 but are restricted by the natural displacement directions in the community in scheme 1. For $\alpha\to\infty$ we are prohibited from making any changes, so again, do not pay much. For system parameters cf. Methods.}
\label{fig:1vs3}
\end{figure}

More specifically, we will first insert a new column $\vec{a}$ (defining how the old species influence the new one) and new row $\vec{b}^T$ (defining how the new species influences the old ones) into the interaction matrix. We then ask how this given $\vec{a}$ should be edited. We do not edit $\vec{b}^T$, not because it is difficult experimentally or analytically, but because it is unnecessary.

After we introduce the new species, the interaction matrix ${\bf A}$ and intrinsic growth rate vector $\vec{r}$, will turn into ${\bf B}$ and $\vec{s}$ defined by
\begin{align}
{\bf B} = \begin{bmatrix}
{\bf A} & \vec{a}+\vec{\epsilon} \\ 
\vec{b}^T & c 
\end{bmatrix}, \quad \vec{s} = \begin{bmatrix}
\vec{r} \\
v
\end{bmatrix},\label{matrices}
\end{align}
where $\vec{\epsilon}$ is our modification to the control species and $(c,v)$ are its self-competition and intrinsic growth. The new interaction matrix ${\bf B}$ and growth rates $\vec{s}$ characterize the new community composed to $N+1$ species. Using block matrix inversion and the fixed point condition we determine how the modification $\vec{\epsilon}$ affects the equilibrium abundance of the introduced species $q$, and that of the original species $\vec{z}$
\begin{equation}
    q = -\frac{\vec{b}\cdot\vec{x} + v}{c-\vec{b}^T{\bf R}(\vec{a}+\vec{\epsilon})},\qquad
    \vec{z} = \vec{x}-q{\bf R}(\vec{a}+\vec{\epsilon}).\label{vec}
\end{equation}

Suppose $\vec{y}$ is the desired target abundances for the original species. We can plug in $\vec{y}$ for $\vec{z}$ above and directly solve for $\vec{\epsilon}^*$ that will get us on target
\begin{equation}
    \vec{\epsilon}^* = c{\bf A}(\vec{y}-\vec{x})/(v+\vec{b}\cdot\vec{y})-\vec{a}.
    \label{eq:exp_a}
\end{equation}

Now we again introduce the cost of making changes to interspecies interactions. We do so by parameterizing the total modification as $\vec{\epsilon}=m\vec{\epsilon}^*$, where $m$ takes a value between $0$ (no modification) and $1$ (reach target). Now that we have a cost the abundances we aim for, $\vec{y}$ will have to be different than the abundances we get $\vec{z}$.

In this case, the modified equilibrium is
\begin{equation*}
    \vec{z}=\vec{x}+\frac{(v+\vec{b}\cdot\vec{x}){\bf R}(\vec{a}+m\vec{\epsilon}^*)}{\vec{b}^T{\bf R}\vec{\epsilon}^*(m_s-m)},\qquad m_s=\frac{c-\vec{b}^T{\bf R}\vec{a}}{\vec{b}^T{\bf R}\vec{\epsilon}^*}.
\end{equation*}
Note that as $m$ approaches $m_s$ we hit the singularity $|\vec{z}|\rightarrow\infty$ as before. We are interested in dialing up the modification magnitude $m$ between 0 and 1 as to minimize the cost function
\begin{equation*}
    L=|\vec{y}-\vec{z}(m)|^2+\alpha m^2 |\vec{\epsilon}^*|^2.
\end{equation*}
As before, the modification cost $\alpha m^2|\vec{\epsilon}^*|$ competes against the distance cost $|\vec{y}-\vec{z}|^2$. We set the derivative to zero and solve for $m^*$
\begin{align}
    \frac{dL}{dm}\bigg|_{m^*}&=-2[\vec{y}-\vec{z}(m^*)]\cdot\frac{d\vec{z}}{dm}\bigg|_{m^*}+2\alpha m^* |\vec{\epsilon}^*|^2=0\nonumber\\   
    \frac{d\vec{z}}{dm}\bigg|_{m^*}&=\frac{(v+\vec{b}\cdot\vec{x}){\bf R}}{\vec{b}^T{\bf R}\vec{\epsilon}^*}\frac{\vec{a}+m_s\vec{\epsilon}^*}{(m_s-m^*)^2}.\nonumber
\end{align}

This is a quartic equation in $m^*$ of the form
\begin{align}
    q_0 + q_1 m^{*} + q_2 m^{*2} + q_3 m^{*3} + q_4 m^{*4} = 0\label{result}
\end{align}
which has an exact analytical solution \cite{weisstein}, given in terms of the coefficients,
\begin{align*}
    q_0 &= 2 |{\bf F}\vec{a}|^2\!+\!(|{\bf F}\vec{\epsilon}^*|^2\!+\!2\vec{a}^T{\bf F}\vec{\epsilon}^*\!-\!2\vec{u}^T{\bf F}\vec{a}) m_s
    \!-\!2\vec{u}^T{\bf F}\vec{\epsilon}^* m_s^2,\\
    q_1 &= 2\vec{a}^T{\bf F}\vec{\epsilon}^*\!+\!|{\bf F}\vec{\epsilon}^*|^2\!+\!2\vec{u}^T{\bf F}\vec{a}\!+\!2\vec{u}^T{\bf F}\vec{\epsilon}^* m_s
    \!+\!2\alpha|\vec{\epsilon}^*|^2 m_s^3,\\
    q_2 &= -6\alpha|\vec{\epsilon}^*|^2 m_s^2,\enskip
    q_3 = 6\alpha|\vec{\epsilon}^*|^2 m_s,\enskip
    q_4 = -2\alpha|\vec{\epsilon}^*|^2,\mbox{ and }\\
    {\bf F} &=(v+\vec{b}\cdot\vec{x}){\bf R}/(\vec{b}^T{\bf R}\vec{\epsilon}^*).
\end{align*}

If the amount of modification is too small to overcome a possible increase in distance due to $\vec{a}$, then we must also reject this solution.


We illustrate our result in Fig.\ref{fig:control_species_vector}. First, we randomly generate a community and a control species (cf. Methods), and then applied an interaction modification to it and introduced it into the community at a time marked by the red dot. We try out two targets, constrained by some small, finite cost. 

In Fig.\ref{fig:control_species_vector}a, we introduce a new species as to eradicate a ``pest'' while leaving all others unaffected. We should emphasize that the control species does not only predate on the pest, but interacts with all species to compensate for the lack of the pest in the final system. In Fig.\ref{fig:control_species_vector}b, we set another target. We introduce a new species as to equalize all abundances. Since in both examples there is a small but finite cost $\alpha$ to making modifications, we approach but do not hit our target abundance. 

In Fig.\ref{fig:control_species_vector}c and \ref{fig:control_species_vector}d we study how well communities of different sizes can be controlled as a function of cost per modification $\alpha$. Fig.\ref{fig:control_species_vector}c shows the fractional reduction in distance from the desired target and  Fig.\ref{fig:control_species_vector}d shows the total cost $\alpha m^2\epsilon^{*2}$ of applying the optimal modification. 

As we see in Fig.\ref{fig:control_species_vector}c and \ref{fig:control_species_vector}d, for very small $\alpha$ values, we end up paying little because it is cheap to make whatever change necessary to get as close as possible to our target. Interestingly, for very large $\alpha$ values, we end up paying little as well. In this case, the cost per change is so high that we are prohibited from making any change. 

Viewing Fig.\ref{fig:control_species_vector}c as a step function, and \ref{fig:control_species_vector}d as a sharp spike at $\alpha_c$, we can say that if the price of modifying the interactions $\alpha$ is cheaper than $\alpha_c$, then we reach our target. Otherwise we do not. 

Interestingly, this critical price point also happens to get us to pay the largest total cost (Fig.\ref{fig:control_species_vector}d). A hypothetical ``interaction modification company'' is best off pricing their services at $\alpha_c$ per modification. If the price is far above, no one will buy their service. If far below, everyone will buy but pay nothing. 

Another interesting economic observation is the non-monotonicity of the peak heights in Fig.\ref{fig:control_species_vector}d as a function of community size. Apparently our hypothetical company profits most from modifying communities that are neither too large, nor too small. 

In Fig.\ref{fig:control_species_vector}c, we see that $\alpha_c$ is smaller for larger communities. This means that it is possible to modify larger communities only with cheap services. This is because control species introduced into large communities must have many interaction values edited. 

{\bf Fine Print: Check if the optimal solution is positive, stable, and close.} So far we solved for optimal perturbations to modify the composition or resilience of a community. However, there are a number of boxes to check before moving forward with these solutions.

First, we must check that the optimal solution does not have negative values for the abundances. This is not because we necessarily get upset when any species go extinct, but because abundances actually hit zero first and remain there first and forever before ever being negative. An optimal solution with negative abundances, regardless of how close it is to the target, will never actually be realized.

Second, we must make sure that our perturbation will not turn an otherwise stable equilibrium point into an unstable one. It is possible for example, that our intervention leads to an unstable equilibrium very near our target, which in practice, does not serve us.

There is an additional check box exclusive to scheme 3, where we are determining how to best edit a particular control species \emph{assuming that the control species is included in the system}. However, it is possible that the best thing to do to an exogenous species in the community, is worse than not including the exogenous species in the first place. Thus, for scheme 3, once we find the optimal modification to the control species, we should check whether not introducing the control species at all leaves us closer to our target. 

\section{Methods}
We verified and illustrated our formulas with simulated communities. We first construct anti-symmetric ($A_{ij}=-A_{ji}$) matrices with normally distributed values with mean $\langle A\rangle=0$ and variance $\sigma_A^2=1$, then reduce all positive interaction values by a factor of $\eta=10$ to account for mass transfer inefficiency (except in Fig.\ref{fig:control_vector_rejection} and Fig.\ref{fig:multi_rejection} where we compare $\eta=10$ and $\eta=1$). We set the diagonal values to $A_{ii}=d=-1$, since most species are self-limiting. Then, we generated normally distributed equilibrium abundances $x_i$ with average $\langle x\rangle=1$ and variance $\sigma_x^2=10^{-2}$, making sure $x_i>0$. Note that $A$ and $x$ determine the growth rate vector according to the coexistent equilibrium condition $\vec{r}=-{\bf A}\vec{x}$. Then we discarded all communities for which $x$ was an unstable equilibrium. To gather statistics we generated $100$ random communities and $100$ random targets for scheme 1, and $1000$ communities and $1000$ random targets for the computationally-cheaper scheme 3. We chose $\zeta=10^{-3}$ for scheme 2 and gather statistics over $1000$ communities. 

For the community size we picked $N=20$ (unless varied in a plot) since communities larger than this tend to be unstable for our parameter values. For the intrinsic growth rate for the control species we always set $v=1$. When determining a target towards which the abundances are to be steered, we pick a random direction, but fix its distance at $|u|=2$.

\section{Discussion}
{\bf Theoretical limits to community control.} Here we have prescribed how to modify the composition and resilience of a community. However in doing so we also established theoretical upper limits to how much a community can be modified at best, i.e. when modification cost poses no constraint. As we see in Fig.\ref{fig:opt_multi} and \ref{fig:prevent_opt_sweep}, the optimal cost function $\mathcal{L}^*$ is bound below, as shown by thin horizontal lines. For control scheme 1, this bound is
\begin{align*}
\mathcal{L}^* = u^2(1-\cos^2\theta_a).
\end{align*} 
For control scheme 1, we showed that hitting our target is possible only if there exist a column of ${\bf A^{-1}}$ that is of the same direction as our aimed displacement $\vec{u}$.

In scheme 2, we showed that it is possible increase the resilience of a community by introducing a very small perturbation to an interaction value. However as the community increases in size, even with no constraint on cost, our ability to stabilize the community becomes severely limited. The best possible performance of control scheme 2 is bounded below by
\begin{align*}
\mathcal{L}^* = \zeta^2\left<x^2\right>\left<\vec{w}_c^2\right>-\zeta^2(\vec{w}_a\cdot\langle \vec{w}\rangle)^2/w_a^2.
\end{align*}

For control scheme 3, we prescribed how to modify exogenous species in order to shift species abundances towards a desirable target. Unlike the first two schemes, we have shown that it is possible to hit our target precisely when not constrained by cost. This relative success stems from modifying multiple interaction elements which allows us to shift the equilibrium of the system in a wide range of directions. 
However, we should also caution that the solution to scheme 3 is to be rejected if it does not check a number of boxes. In our numerical tests, we have observed that if the control species is chosen without care (we randomly generate the natural interactions of the control species, as opposed to determining the ``best'' control species from available options) the probability of rejection can be rather significant. We show in the appendix the rejection rates for various reasons, for various costs and community sizes, when the natural interactions of the control species with others are randomly generated.

What if we were allowed to modify not all of the interactions of a control species with others, but only some of them? In this case we would substitute 0 for the unchangeable components of $\vec{\epsilon}$ in Eqn.(\ref{vec}), and substitute our target $\vec{z}\to \vec{y}$ and hope that these overdetermined set of equations have a solution for $\vec{y}$. That is to say, we can shift the equilibrium exactly onto $\vec{y}$, only when our target vector $\vec{u}$ can be written as a linear combination of $\vec{w}_i$, where $i$ indexes the changeable components of $\vec{\epsilon}$.

We should be clear that our framework has many restrictions, as it hinges on well-mixedness, quasi-equilibrium, negligible demographic noise, trivial functional responses, and exclusively two-body interactions. Nevertheless, our results are transparent, analytical and illustrative, and should therefore serve as a valuable guideline for experimental, natural, agricultural and industrial systems that approximately satisfy these properties. It should serve also as a comparative benchmark and a limiting case for more complicated mathematical models of population dynamics that do not admit analytical solutions as easily.

\newpage

{\bf Acknowledgement.} This material is based upon work supported by the Defense Advanced Research Projects Agency under Contract No. HR0011-16-C-0062

\appendix
\section{Appendix: A detailed analysis of rejected solutions}

\begin{figure}
\centering
\includegraphics[width=\linewidth]{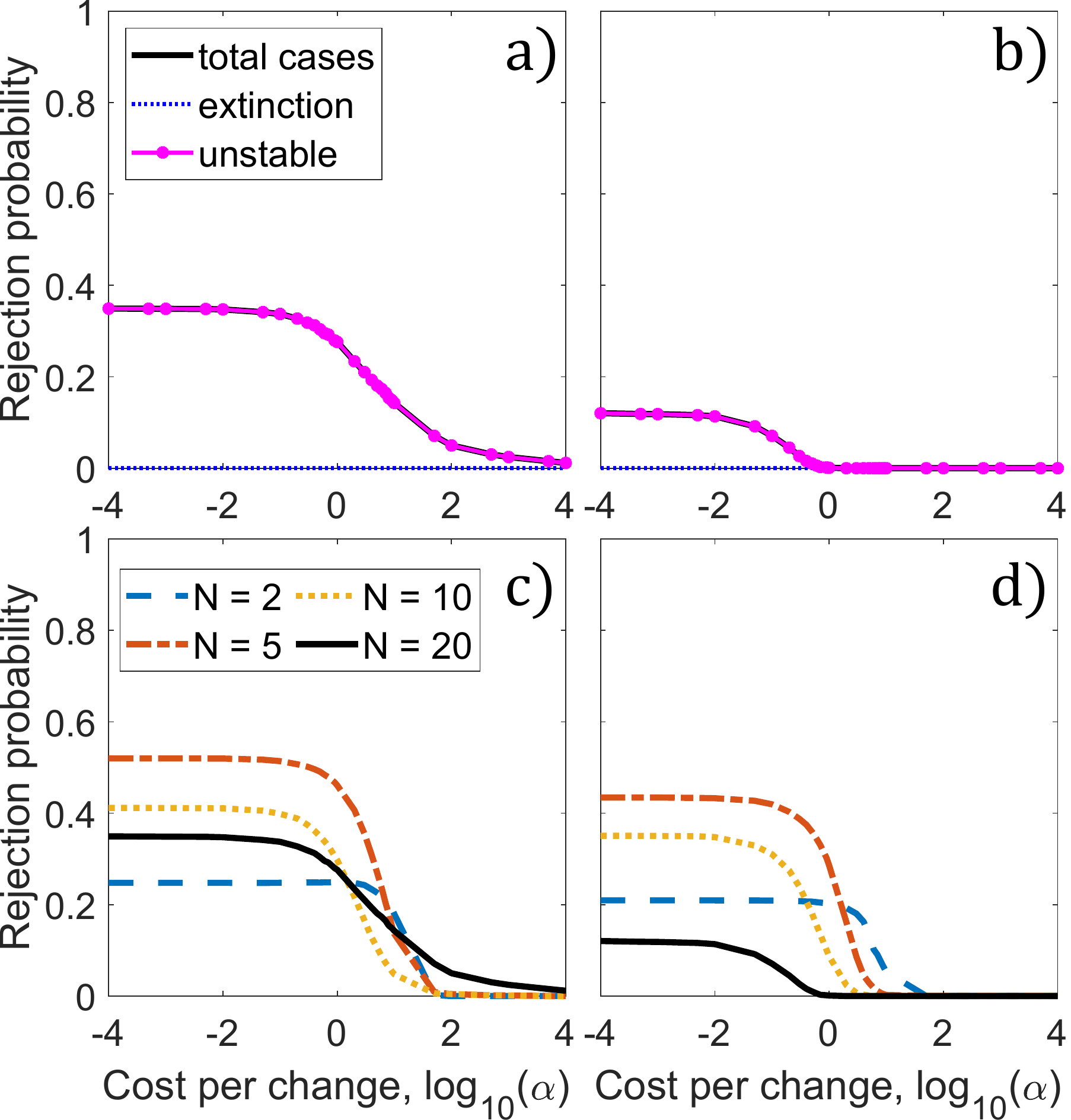}
\caption{{\bf  Control Scheme 1: Rejecting the optimal solution.} {\bf Top Row:} The optimal modification in a single interaction can be rejected when it destabilizes the community. The efficiency of mass transfer between prey to predator affects the stability of the communities. When the efficiency is low ({\bf left column}, $\eta=
10$) there is a higher chance that strong interactions destabilizes the ecosystem in comparison to highly efficient communities ({\bf right coloumn}, $\eta=1$). {\bf Bottom Row}: Total rejection probability for different community sizes. For system parameters cf. Methods.}
\label{fig:multi_rejection}
\end{figure}

\begin{figure}
\centering
\includegraphics[width=\linewidth]{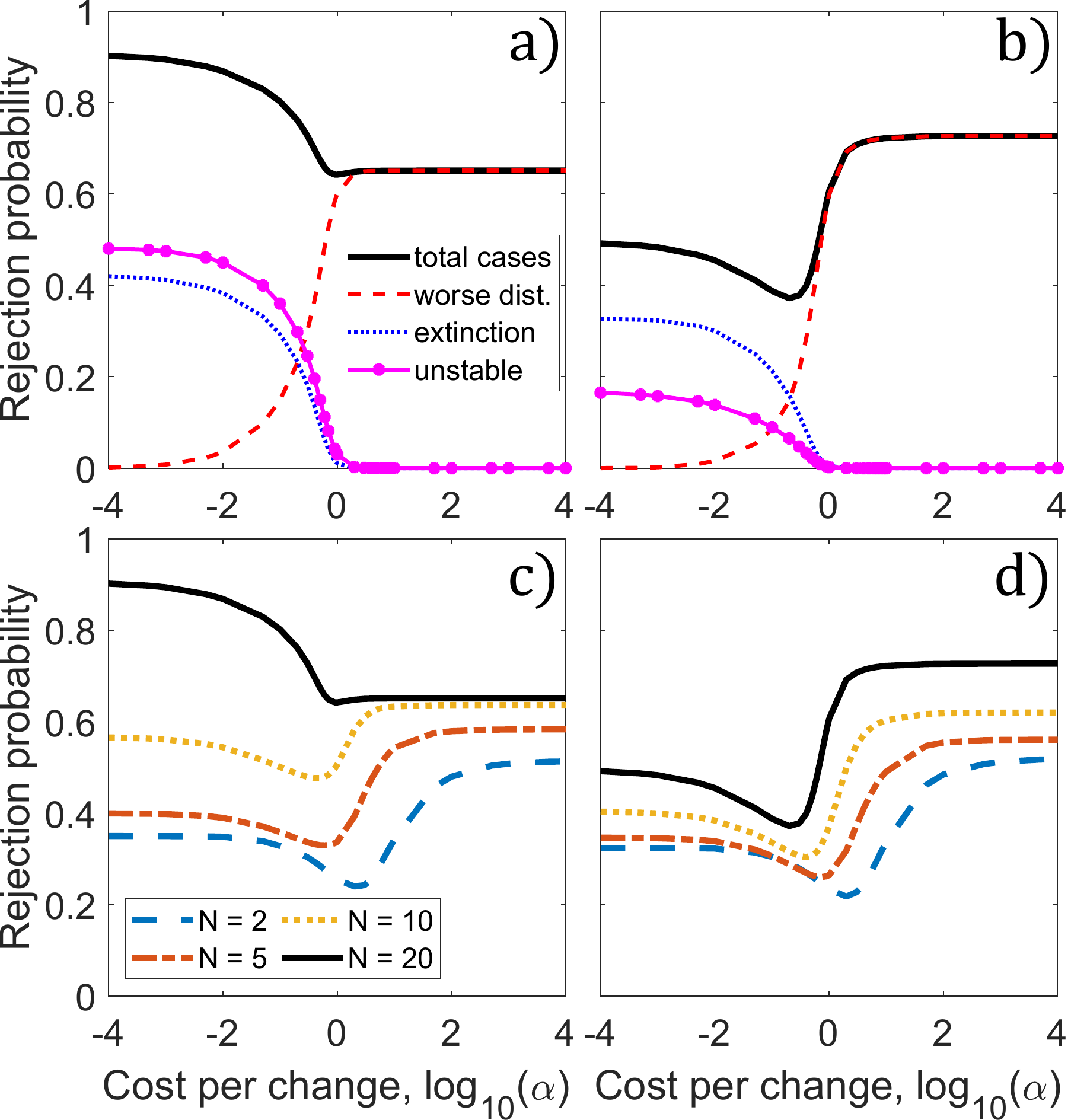}
\caption{{\bf  Control Scheme 3: Rejecting the optimal solution.} {\bf Top Row:} Reasons for rejection. For a given community structure and a target, a randomly chosen control species (i.e. randomly generated $\vec{a}$) will often not check the boxes defined in the main text, and should not be used as a control species. The black line shows how often this will happen, as a function of log modification cost. For low $\alpha$ the optimal solution is rejected because the large modifications, and thus large interaction values, drive some species (usually the exogenous species itself) to extinction. For high $\alpha$, the natural interaction values of the control species will largely remain unchanged. Since these interaction values are randomly generated, they typically do not get us close to our target, and thus are rejected. Here $N=20$. {\bf Bottom Row}: Total rejection probability for different community sizes. Smaller communities are easier to control. The optimal solution for a randomly generated control species is far likely to pass the stability, extinction and distance tests for smaller communities. The efficiency of mass transfer between prey to predator affects the stability of the communities. When the efficiency is low ({\bf left column}, $\eta=
10$) there is a higher chance that strong interactions destabilizes the ecosystem in comparison to highly efficient communities ({\bf right coloumn}, $\eta=1$). For other system parameters cf. Methods.}
\label{fig:control_vector_rejection}
\end{figure}

In control scheme 1, the probability of rejecting the optimal modification in the single best interaction element shown in Fig.\ref{fig:multi_rejection}a is high only in the ($\alpha\ll\alpha_c$) limit. When the modification cost is low, the introduced change can be large enough to potentially destabilize the community. Rejection probabilities drops to zero as the modification cost increases. 

In control scheme 2, the rejection probabilities are less than $1\%$ for values sampled in Fig.\ref{fig:prevent_opt_sweep}. Since this is quite insignificant, we do not analyze these rejections in depth.

In control scheme 3, the rejections in the ($\alpha\ll\alpha_c)$ limit are due to introducing strong interactions which will destabilize the community or cause some species (typically the introduced control species itself) to go extinct. This is not a defect of our control scheme but rather because it is more difficult to safely manipulate communities that are close to the edge of instability. Compare for example, the black lines in Fig.\ref{fig:control_vector_rejection}a and b, corresponding to communities with higher (more stable) and lower (less stable) mass conversion efficiency. In Fig.\ref{fig:control_vector_rejection}c and d we also see that smaller communities also lead to less rejections compared to larger ones since, as is well known \cite{May}, smaller communities are more stable to begin with.

In the ($\alpha\gg\alpha_c)$ limit, the reason for the rejections can be understood as follows. Since in our numerical tests the natural interaction values of the control species were chosen randomly, when unmodified, the new species tend to shift the original equilibrium to some random direction; sometimes closer to our target, but typically further away. However, if the price $\alpha$ is high enough, the optimal $\epsilon$ turns out to not large enough to overcome this random shift to get us closer to our target. 

Thus, the moral is that one should not carelessly pick ``any'' species (as we do in our numerical tests), especially if the modification cost is high. Instead, one should consider a number of options, obtain the optimal solution for each candidate control species using Eqn.(\ref{result}), verify that it checks the negativity, stability and closeness checkboxes, and only then use the species with the least $\mathcal{L}^*$ with the appropriate interaction edits $\vec{\epsilon}$. 

\end{document}